\documentclass[]{spie}  

\newcommand{\umux}{$\mu$mux\,}
 
\usepackage{amsmath,amsfonts,amssymb}
\usepackage{graphicx}
\usepackage[colorlinks=true, allcolors=blue]{hyperref}

\usepackage{subcaption}
\usepackage[export]{adjustbox}
\usepackage{xcolor}

\title{SPT-3G+: A Cosmic Microwave Background Experiment for the South Pole Telescope}

\author[a,b]{T.~Natoli}
\author[c,d]{Z.~Ahmed}
\author[a,b,e]{H.~Athreya}
\author[f]{J.~E.~Austermann}
\author[f]{K.~Bae}
\author[g,e]{A.~Bapat}
\author[h]{D.~R.~Barron}
\author[i]{P.~S.~Barry}
\author[e,b,a]{A.~N.~Bender}
\author[j,b,a]{B.~A.~Benson}
\author[e,b,a]{L.~E.~Bleem}
\author[b,k,e,a,l]{J.~E.~Carlstrom}
\author[e]{T.~W.~Cecil}
\author[b,e,a]{C.~L.~Chang}
\author[d]{S.~Cisneros}
\author[m]{A.~Coerver}
\author[e]{J.~Cornelison}
\author[a,b]{T.~M.~Crawford}
\author[a]{R.~Datta}
\author[a,b]{K.~R.~Dibert}
\author[h]{W.~Dominguez}
\author[f]{S.~M.~Duff}
\author[k,b]{K.~Fichman}
\author[n,o]{J.~P.~Filippini}
\author[p]{L.~Gades}
\author[q,b]{P.~A.~Gallardo}
\author[r]{S.~Galli}
\author[s,t]{N.~W.~Halverson}
\author[o]{Q.~Hao}
\author[d]{S.~Henderson}
\author[d]{R.~Herbst}
\author[m]{W.~L.~Holzapfel}
\author[k,b]{A.~Hryciuk}
\author[f]{J.~Hubmayr}
\author[f,t]{D.~Jones}
\author[k]{V.~Kabra}
\author[u]{K.~S.~Karkare}
\author[v]{C.~King}
\author[f,t]{M.~A.~Koc}
\author[b,a,k]{A.~M.~Kofman}
\author[f,w]{R.~A.~Lew}
\author[f]{M.~J.~Link}
\author[f]{T.~J.~Lucas}
\author[k]{A.~Mangu}
\author[a,b]{E.~S.~Martsen}
\author[b,k,a]{J.~J.~McMahon}
\author[x]{J.~Montgomery}
\author[v]{J.~M.~Nagy}
\author[j]{H.~Nguyen}
\author[y,z]{V.~Novosad}
\author[aa]{S.~Padin}
\author[d]{T.~Pinsonneault-Marotte}
\author[bb,cc]{S.~Raghunathan}
\author[a,b]{A.~S.~Rahlin}
\author[dd]{C.~L.~Reichardt}
\author[d]{L.~Ruckman}
\author[v]{J.~E.~Ruhl}
\author[v]{M.~S.~Sarwar}
\author[j,a]{S.~Simon}
\author[f]{R.~Singh}
\author[ee,j]{J.~A.~Sobrin}
\author[ff]{A.~A.~Stark}
\author[v]{Q.~Taylor}
\author[i]{C.~Tucker}
\author[f]{J.~Ullom}
\author[f]{J.~Van Lanen}
\author[o,n,cc]{J.~D.~Vieira}
\author[b,a,l,k]{A.~G.~Vieregg}
\author[f]{M.~R.~Vissers}
\author[e]{G.~Wang}
\author[aa]{W.~L.~K.~Wu}
\author[e]{V.~Yefremenko}
\author[gg]{E.~Yilmaz}
\author[e,b]{C.~Yu}
\author[a]{J.~Zivick}

\affil[a]{Department of Astronomy and Astrophysics, University of Chicago, 5640 South Ellis Avenue, Chicago, IL, 60637, USA}
\affil[b]{Kavli Institute for Cosmological Physics, University of Chicago, 5640 South Ellis Avenue, Chicago, IL, 60637, USA}
\affil[c]{Kavli Institute for Particle Astrophysics and Cosmology, Stanford University, 452 Lomita Mall, Stanford, CA, 94305, USA}
\affil[d]{SLAC National Accelerator Laboratory, 2575 Sand Hill Road, Menlo Park, CA, 94025, USA}
\affil[e]{High-Energy Physics Division, Argonne National Laboratory, 9700 South Cass Avenue, Lemont, IL, 60439, USA}
\affil[f]{Quantum Sensors Division, National Institute of Standards and Technology, 325 Broadway, Boulder, CO, 80305, USA}
\affil[g]{Pritzker School of Molecular Engineering, University of Chicago, 5640 S Ellis Avenue, Chicago, IL 606037, USA}
\affil[h]{Department of Physics and Astronomy, University of New Mexico, Albuquerque, NM, 87131, USA}
\affil[i]{School of Physics and Astronomy, Cardiff University, Cardiff CF24 3YB, United Kingdom}
\affil[j]{Fermi National Accelerator Laboratory, MS209, P.O. Box 500, Batavia, IL, 60510, USA}
\affil[k]{Department of Physics, University of Chicago, 5640 South Ellis Avenue, Chicago, IL, 60637, USA}
\affil[l]{Enrico Fermi Institute, University of Chicago, 5640 South Ellis Avenue, Chicago, IL, 60637, USA}
\affil[m]{Department of Physics, University of California, Berkeley, CA, 94720, USA}
\affil[n]{Department of Physics, University of Illinois Urbana-Champaign, 1110 West Green Street, Urbana, IL, 61801, USA}
\affil[o]{Department of Astronomy, University of Illinois Urbana-Champaign, 1002 West Green Street, Urbana, IL, 61801, USA}
\affil[p]{X-ray Science Division, Argonne National Laboratory, 9700 South Cass Avenue, Lemont, IL, 60439, USA}
\affil[q]{Department of Physics \& Astronomy, University of Pennsylvania, 209 S. 33rd Street, Philadelphia, PA 19064, USA}
\affil[r]{Sorbonne Universit\'e, CNRS, UMR 7095, Institut d'Astrophysique de Paris, 98 bis bd Arago, 75014 Paris, France}
\affil[s]{Department of Astrophysical and Planetary Sciences, University of Colorado, Boulder, CO, 80309, USA}
\affil[t]{Department of Physics, University of Colorado, Boulder, CO, 80309, USA}
\affil[u]{Department of Physics, Boston University, Boston, MA 02215, USA}
\affil[v]{Department of Physics, Case Western Reserve University, Cleveland, OH, 44106, USA}
\affil[w]{Theiss Research, La Jolla, CA, 92037, USA}
\affil[x]{t0.technology, 2200-300 Rue Leo-Pariseau, Montreal, Q.C., H2X 4B3, Canada}
\affil[y]{Materials Sciences Division, Argonne National Laboratory, 9700 South Cass Avenue, Lemont, IL, 60439, USA}
\affil[z]{Institute of Multidisciplinary Research for Advanced Materials, Tohoku University, Sendai, 980-8577, Japan}
\affil[aa]{California Institute of Technology, 1200 East California Boulevard, Pasadena, CA, 91125, USA}
\affil[bb]{Department of Physics \& Astronomy, University of California, One Shields Avenue, Davis, CA 95616, USA}
\affil[cc]{Center for AstroPhysical Surveys, National Center for Supercomputing Applications, Urbana, IL, 61801, USA}
\affil[dd]{School of Physics, University of Melbourne, Parkville, VIC 3010, Australia}
\affil[ee]{Department of Physics, Villanova University, 800 E Lancaster Ave., Villanova, PA 19085, USA}
\affil[ff]{Center for Astrophysics \textbar{} Harvard \& Smithsonian, 60 Garden Street, Cambridge, MA, 02138, USA}
\affil[gg]{Department of Electrical and Computer Engineering, University of Illinois Urbana-Champaign, 306 N Wright St, Urbana, IL 61801, USA}

\authorinfo{Corresponding author: Tyler Natoli, tnatoli@uchicago.edu}

\newcommand{\sptthreeg}{SPT-3G}
\newcommand{\sptthreegplus}{SPT-3G$+$} 
\newcommand{\sqdeg}{{\rm deg}^{2}}
\newcommand{\fieldsize}{800}

\newcommand{\snr}{S/N}
\newcommand{\mvir}{M_{500c}}
\newcommand{\msol}{M_{\odot}}
\newcommand{\nzero}{N_{L}^{0}}

\begin{document} 
\maketitle

\begin{abstract}
SPT-3G+ is the next survey receiver planned to be installed in early 2029 on the 10-meter South Pole Telescope (SPT). 
This new receiver will feature 6,020 polarization-sensitive dichroic pixels with transition-edge sensors observing in frequency bands centered at 90 GHz and 150 GHz.
The 24,080 detectors in the \sptthreegplus{} receiver will be cooled to 100 mK by a dilution refrigerator and read out using microwave SQUID multiplexing. 
The optical design of the receiver enables a 4~degree diameter field of view, which is broken up into 14 individual optics tubes each containing cryogenic alumina, silicon, and nylon lenses. 
These technology choices will allow the SPT-3G+ receiver to improve on the mapping speed of the currently operating SPT-3G receiver by nearly an order of magnitude.
Once deployed, the SPT-3G+ receiver will observe for 6-years an area overlapping with the BICEP survey to achieve a combined (90 GHz and 150 GHz) CMB map depth of 0.5 uK-arcmin. 
Data from these observations will be used to create unprecedentedly deep CMB lensing maps, discover new galaxy clusters, and detect astrophysical transients. 
The lensing map produced by SPT-3G+ will be used to remove or “delens”  foreground B modes, where large-scale structure gravitationally lenses the CMB and converts E modes into B-mode polarization, 
with the goal of revealing inflationary B modes. 
Together with data from the BICEP Array as part of the South Pole Observatory, SPT-3G+ data will be used to constrain the tensor-to-scalar ratio $r$ with a goal of achieving a measurement of $\sigma(r) = 0.001$.
\end{abstract}


\section{INTRODUCTION}
\label{sec:intro}  

The South Pole is one of the best sites in the world for mm-wavelength observations due to its low precipitable water vapor, high altitude, and stable atmosphere \cite{bussmann05}. 
In 2006, the 10-meter mm-wavelength South Pole Telescope (SPT) was constructed at the NSF Amundsen-Scott South Pole Station to take advantage of these favorable conditions for cosmic microwave background (CMB) observations \cite{ruhl04,carlstrom11}. 
There have been three CMB survey instruments installed on the SPT to date: SPT-SZ, SPTpol, and \sptthreeg. 

The first instrument on the SPT, SPT-SZ, was installed from 2007 to 2011 and observed a $2,500~\sqdeg$ patch of sky to a white-noise level of 36, 16, and 62 $\mu$K-arcmin at 95, 150, and 220~GHz, respectively, in CMB temperature anisotropy units \cite{bleem15b}.
The second instrument, SPTpol, was installed from 2012 to 2016 and observed at 95 GHz and 150 GHz with polarization sensitivity \cite{austermann12}. 
The SPTpol instrument performed a $500~\sqdeg$ survey with a final temperature white-noise level at 95 and 150~GHz of 5.9$ \mu$K-arcmin and 13.5 $\mu$K-arcmin \cite{chou25}. 
In 2017, the \sptthreeg{} instrument was installed on the SPT along with new ambient-temperature secondary and tertiary mirrors.
This upgraded system allows the tri-chroic polarization-sensitive SPT-3G focal plane to observe with a 1.88 deg diameter field of view \cite{sobrin18}.
\sptthreeg{} is still installed on the SPT and continuing a $1500~\sqdeg$ main survey, which will achieve an expected temperature depth of 2.2, 1.9, 6.7 $\mu$K-armcin at 95, 150, and 220~GHz, respectively, with data through the end of 2028 \cite{prabhu24}. 

The next instrument to be installed on the SPT is \sptthreegplus, which will have a mapping speed nearly an order of magnitude larger than the currently installed \sptthreeg{}. 
This new \sptthreegplus\, instrument will feature a 24,080 detector focal plane and start observations in 2029.  \sptthreegplus\, will be used on a 6-year survey of  $\fieldsize~\sqdeg$, aiming to achieve a temperature depth of 0.7~$\mu$K-arcmin at both 90 and 150~GHz. 
In this paper we will discuss the scientific goals, performance capabilities, and design specifics for the \sptthreegplus instrument. 

\section{Science Goals}
\label{science}

Cosmological constraints from new measurements of the CMB are largely driven by making high precision measurements of the polarization anisotropy of the CMB 
\cite{planck18-1, louis25,chou25,quan26}.
The polarization signal of CMB maps can be separated into two distinct polarization patterns; curl-free E modes and divergence-free B modes.
E modes are created by Thomson scattering near quadrupole anisotropies during the formation of the CMB and can improve $\Lambda$CDM parameter constraints when combined with CMB temperature data \cite{galli14}. 
B-mode signals can be created by inflationary gravitational waves \cite{seljak97}, and by gravitational lensing of E modes as light from the CMB interacts with matter while free-streaming through the universe \cite{zaldarriaga98}. 
To improve constraints on the inflationary signal, the `lensing B modes' can be removed through a process called `delensing', described further in \ref{cosmic_inflation}.  
The inflationary B-mode signal is expected to peak at spatial scales around a degree, with measurements in this range able to constrain the tensor-to-scalar ratio, $r$, which can measure measure the energy scale of inflation and rule out specific models of inflation\cite{kamionkowski15}.

No experiment has detected inflationary B modes yet, with experiments only placing upper limits on $r$ so far. 
The most stringent constraint on $r$ comes from the BICEP/Keck experiment \cite{bicepkeck21c}.
BICEP/Keck represents a series of experiments using small aperture telescopes at the South Pole with an angular resolution of $\sim$1~degree\cite{bicep14b}.
These small aperture experiments lack the resolution needed to measure lensing B-modes on their own.
To create an estimate of the lensing B mode pattern and remove its contamination of the $r$ estimate, a tracer of the CMB lensing potential must be measured. 
For this, a large aperture telescope, like the SPT, is needed to produce a high signal-to-noise reconstruction of the CMB lensing potential with sufficient resolution for delensing.  

For SPT to most effectively remove the lensing B mode signal from BICEP maps, it needs to make the lowest-noise CMB maps possible over the same survey field as BICEP. 
Starting in 2029, \sptthreegplus{} will begin a six year survey of an $\fieldsize~\sqdeg$ patch of sky centered on the BICEP survey field. 
The overlapping surveys will allow the high-resolution \sptthreegplus\, lensing map, discussed in Section \ref{lensing}, to be used to remove the lensing B-mode signal in BICEP maps, as discussed in Section \ref{cosmic_inflation}. 

Although the primary measurement goal of \sptthreegplus{} is to measure CMB lensing with sufficient sensitivity to delens the BICEP maps, the deep $\sim$1~arcmin resolution CMB maps it produces will also be used to extract other compelling science.
\sptthreegplus{} will make new precision measurements of the thermal Sunyaev-Zel'dovich effect (tSZ) \cite{sunyaev72}, more than doubling the number of clusters detected per $\sqdeg$ compared to \sptthreeg, which will unlock new measurements of the clustering of clusters, and probe even earlier epochs of cluster formation.
\sptthreegplus{} will also probe new regimes of the time-variable millimeter-wavelength sky, due to its factor of several improvement in instantaneous sensitivity compared to \sptthreeg, and $\sim$daily observations of the full $\fieldsize~\sqdeg$ field, allowing the monitoring and discovery of new populations of transient sources \cite{guns21, chichura22, hood23, tandoi24}.

\subsection{Gravitational Lensing Map}
\label{lensing}

\begin{figure}[ht]
\centering
\includegraphics[width=0.8\textwidth, keepaspectratio]{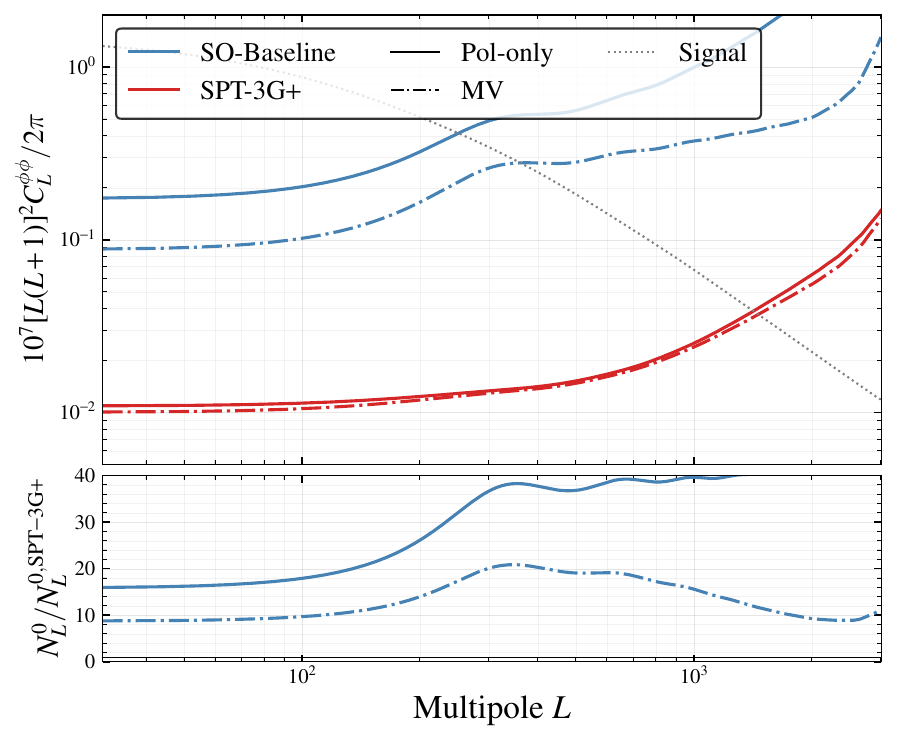}
\caption
{
The anticipated lensing noise per mode, $\nzero$ for optimal lensing reconstruction: polarization-only (solid) and the minimum-variance (MV; dash-dotted) combination of all lensing estimators, shown for the \sptthreegplus{} (red) and SO (blue) experiments. 
The fiducial lensing power spectrum is shown as a gray dotted line. 
As illustrated in the bottom panel, the lensing reconstruction noise of \sptthreegplus{} is significantly lower than that of the SO Baseline\cite{abitbol25} sensitivity across a wide range of scales. 
In particular, on the large angular scales most relevant for delensing to facilitate the primordial B-mode searches with SPO, the deep polarization-only data from \sptthreegplus{} will enable lensing maps with a $\snr$ approximately $9\times$ higher than SO.
}
\label{fig_lensing_noise}
\end{figure}

The \sptthreegplus{} instrument will make deep temperature and polarization maps of the CMB that will be used to reconstruct the gravitational lensing field with high signal-to-noise. 
The resulting lensing map will be sample-variance dominated down to arcminute scales ($L \sim 1000$). 
Figure~\ref{fig_lensing_noise} shows the lensing reconstruction noise per mode, $\nzero$, for \sptthreegplus{} (red) and Simons Observatory (blue) with the `baseline' sensitivity  listed in Abitbol et al.\cite{abitbol25}. 
The dash-dotted curves correspond to lensing reconstruction using both temperature and polarization data, while the solid curves show the performance obtained using polarization-only measurements.
The lensing noise is computed using the iterative algorithm in Smith et al.\cite{smith2012} assuming $\ell_{\rm max} = 3500$ for temperature and $\ell_{\rm max} = 4000$ for polarization. 
The large aperture telescope in the Simons Observatory is conducting a shallower but much wider $\sim$$25,000~\sqdeg$ survey compared to the $\fieldsize~\sqdeg$ field size of \sptthreegplus{}. 
Compared to the \sptthreegplus{} measurements, the Simons Observatory lensing maps result in a relatively smaller improvement in $r$ constraints from delensing, however is more optimal for the overall cosmological constraining power from the CMB temperature and polarization power spectra \cite{abitbol25}.

Because astrophysical foregrounds are largely unpolarized \cite{datta19, gupta19}, polarization-only lensing reconstruction is significantly less susceptible to foreground-induced systematic biases than reconstructions that rely on CMB temperature data \cite{vanengelen12, namikawa13, madhavacheril18, lembo22, raghunathan23}. 
As shown in Figure \ref{fig_lensing_noise}, on scales $L < 500$, the lensing $\snr$ per mode achieved by \sptthreegplus{} is a factor of  $\sim$$9$ higher than SO.
The $\snr$ per mode is a more important metric than lensing bandpower errors when removing lensing-induced B-mode polarization to measure $r$, as discussed in Section \ref{cosmic_inflation}.

\subsection{Inflationary B Modes}
\label{cosmic_inflation}

The South Pole Observatory (SPO) is a coordinated observing and analysis program by the SPT and the BICEP collaborations, which will work to jointly analyze CMB maps taken from the South Pole. 
The main goal of the SPO is to detect inflationary B modes and measure the tensor-to-scalar ratio, $r$. 
At the current sensitivity of the BICEP/Keck data, the constraints on $r$ can be significantly improved by removing ``lensing B modes", which are B modes created from gravitationally lensing E modes, through a process called ``delensing'' \cite{bicepkeckspt21}.

As discussed in Section \ref{science}, a large aperture telescope is required to produce a lensing map with sufficient angular resolution to resolve and delens the lensing B modes. 
Since SPT and BICEP are at the same South Pole site, they can continuously observe the same patch of sky. 
The \sptthreegplus{} $\fieldsize~\sqdeg$ survey field was chosen to fully overlap with the patch of sky observed by BICEP/Keck. 
This overlap in fields will allow SPO to use the high-fidelity lensing map produced by \sptthreegplus{} to remove lensing B modes from the low noise BICEP/Keck maps.
With these lensing B modes removed, SPO will place a limit of $\sigma(r) \sim 0.001$ using data through 2034. 
To put this in context, the current best limit on $\sigma(r)$ is from BICEP/Keck at 0.009 \cite{bicepkeck21c}, and the predicted $\sigma(r)$ for the Simons Observatory with 10 years of data is 0.0014 \cite{abitbol25}.

\subsection{Galaxy Clusters}

Galaxy cluster samples selected using the tSZ effect are mass-limited samples. 
With its deep, high-resolution maps, the \sptthreegplus{} survey will detect clusters at $\snr=5$ ($>99.5\%$ purity threshold) with cluster masses of \mbox{$\mvir \ge 8 \times 10^{13} \msol$} at the epoch of cosmic noon ($z \approx 2$) \cite{raghunathan22b}. 
The number of clusters detected per square degree by \sptthreegplus{} will increase by a factor of $2$ compared to \sptthreeg.
This increase in cluster density will better enable studies of the clustering of galaxy clusters\cite{cromer19}, and the abundance and formation history of clusters around cosmic noon \cite{cmbs4collab19, raghunathan22b}.

\begin{figure}[ht]
\centering
   \includegraphics[width=0.7\textwidth, keepaspectratio]{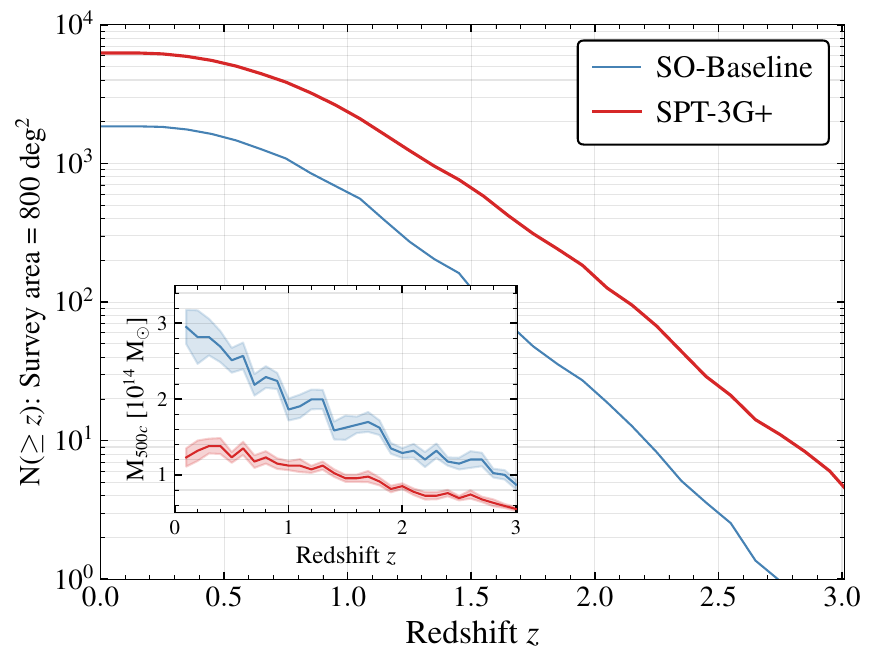}
   \caption
   {The expected number of SZ-selected clusters with $\snr \ge 5$ in an $\fieldsize~\sqdeg$ survey footprint is shown, with \sptthreegplus{} in red and Simons Observatory (SO, baseline sensitivity)\cite{abitbol25} in blue. 
   While we’ve chosen to compare cluster counts across an $\fieldsize~\sqdeg$ patch, note that SO plans to observe $\sim$25,000 $\sqdeg$ to the same noise level. 
   However, within comparable areas of sky, \sptthreegplus{} will detect 4x more clusters with $\snr>5$ than SO. 
   The cluster mass, $\mvir$, that can be detected at $\snr \ge 5$ for each experiment across redshift is  illustrated in the inset, highlighting the depth of the \sptthreegplus{} survey. 
   }
\label{fig_cluster_counts}
\end{figure}

Figure~\ref{fig_cluster_counts} shows the expected number of tSZ-selected clusters with $\snr \ge 5$ for \sptthreegplus{} and the Simons Observatory with baseline sensitivity\cite{abitbol25}.
Compared to the Simon Observatory's wider survey strategy, which will yield larger absolute cluster samples, the smaller field size of the \sptthreegplus{} survey will enable the largest galaxy cluster densities to date and detect lower-mass, higher redshift clusters.
Across all redshifts, the mass threshold of the \sptthreegplus{} survey will be a factor of $2\times$ lower relative to the Simons Observatory (baseline sensitivity)\cite{abitbol25}.

\section{Instrument}

The new \sptthreegplus{} instrument is designed to integrate the latest technology into the South Pole Telescope to achieve an order of magnitude faster mapping speed than the \sptthreeg{} instrument.
This improvement in mapping speed is the product of a larger field of view, higher detector packing density, better optical efficiency, and lower readout noise.
To model the mapping speed of the \sptthreegplus{} instrument, the noise equivalent temperature (NET) of each observing band was calculated with the publicly available jbolo\footnote{https://github.com/JohnRuhl/jbolo}  which is based on BoloCalc \cite{hill18}.
This code takes into account calculations of the losses and in-band emission of each optical element as a function of frequency, as well as the detector properties, including photon, phonon, Johnson, and readout noise contributions.

\subsection{Optical Design} \label{Optical}

\begin{figure}
    \centering
    \begin{minipage}{0.45\textwidth}
        \centering
        \includegraphics[width=1.1\textwidth]{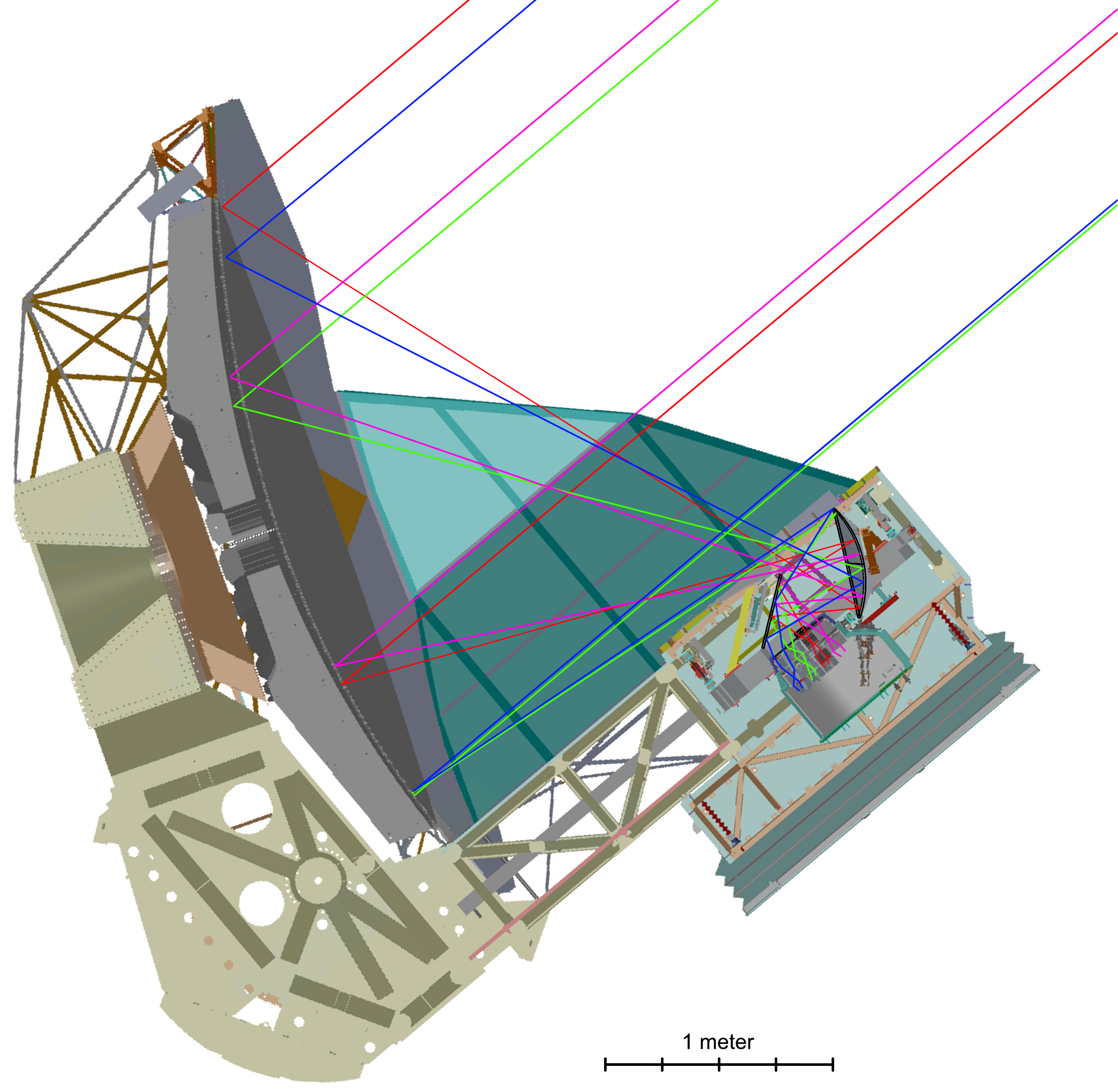}
    \end{minipage}\hfill
    \begin{minipage}{0.45\textwidth}
        \centering
        \includegraphics[width=0.9\textwidth]{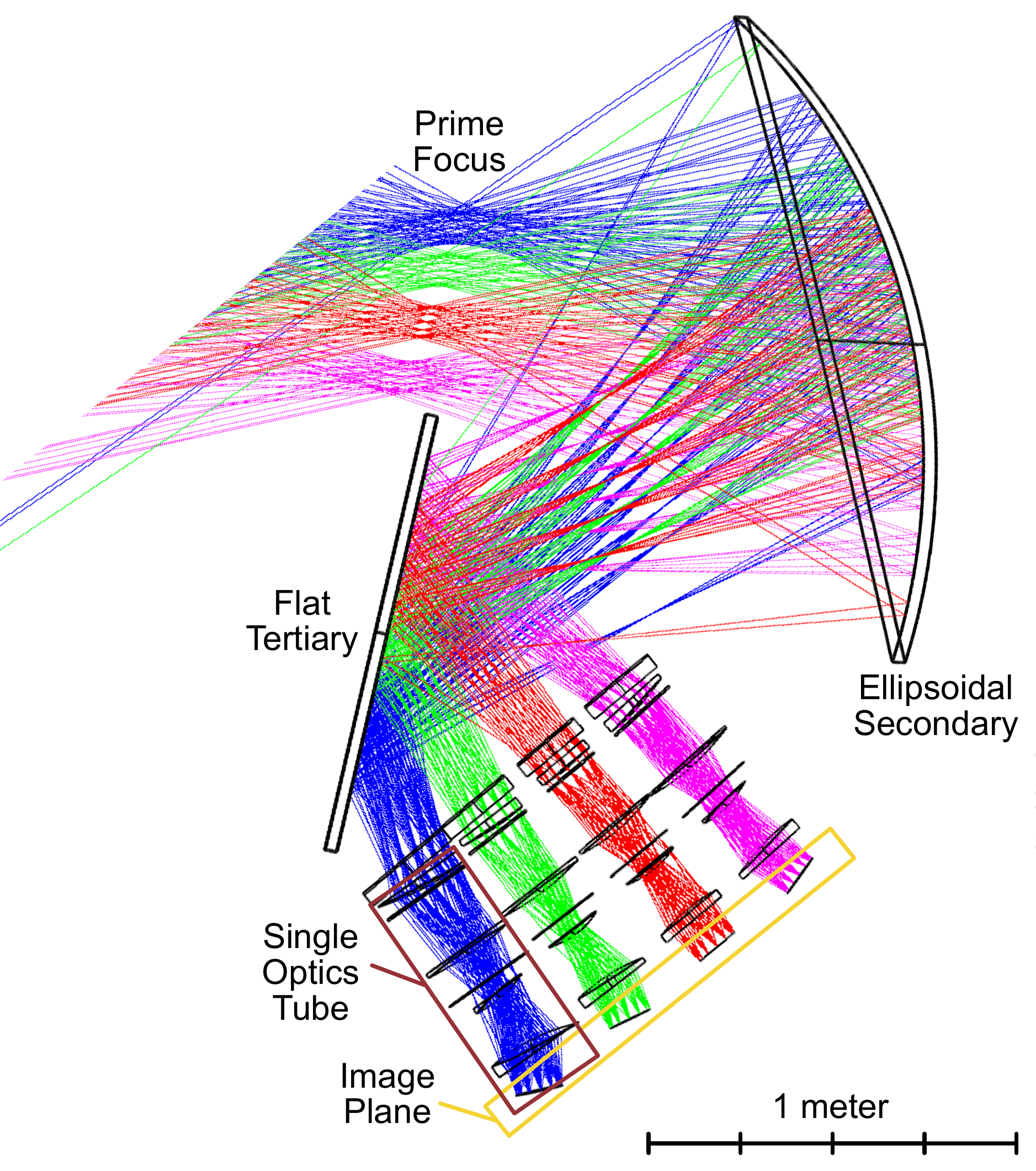}
    \end{minipage}
    \caption{\textbf{Left}: A cross-sectional view of the SPT with the \sptthreegplus{} cryostat inside the telescope cabin. 
    The colored lines show rays traced from the four optics tubes along the centerline of the telescope. 
    The telescope is shown looking at an elevation of 40 deg, which is the elevation where the pulse tube cryocoolers are oriented vertically.
    \textbf{Right}: A detailed view of the ray tracing inside the telescope cabin. }
    \label{fig:SPT_optics}
\end{figure}

The SPT is an off-axis Gregorian telescope with a 10~meter diameter primary mirror that is surrounded by a $>$1.5~m extension shield to prevent scattered light from the ground from entering the receiver\cite{padin08,austermann12}. 
As seen in Figure \ref{fig:SPT_optics}, light reflected from the SPT primary mirror is directed into the telescope receiver cabin where it is again reflected off ambient-temperature secondary and tertiary mirrors.
The \sptthreegplus{} instrument will use the same 1.7 meter diameter ellipsoidal secondary as \sptthreeg \cite{sobrin18,stark18}, but will use a new 1.1 meter diameter flat tertiary mirror .
Following reflection off the tertiary mirror, light enters the cryogenic receiver through a large, monolithic high-density polyethylene (HDPE) vacuum window spanning the entire active optical area.

\begin{figure} [ht]
   \begin{center}
   \begin{tabular}{c} 
   \includegraphics[height=10cm]{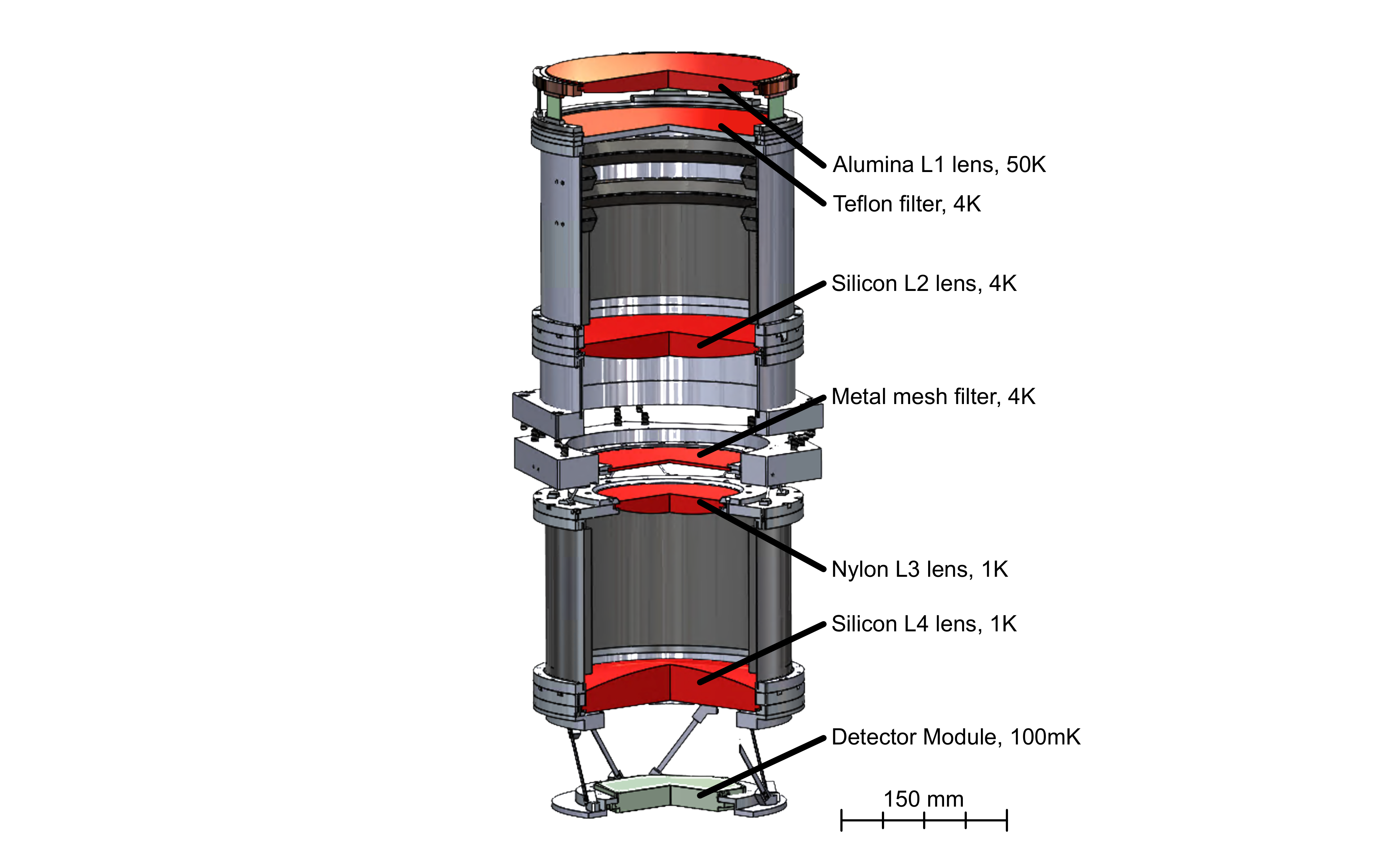}
   \end{tabular}
   \end{center}
   \caption[example] 
   { \label{fig:optics_tube} 
    An example of one of the \sptthreegplus{} optics tubes with the temperature of optical elements labeled. 
    Across the 14 optics tubes in \sptthreegplus there are 7 different optical tube designs with each design having unique lens shapes. 
}
   \end{figure}

While previous receivers on the SPT have utilized a single chain of refractive-optical elements and infrared-blocking filters, the \sptthreegplus{} receiver will have a different optics chain for each of the 14 detector wafers. 
Each optics chain is referred to as an `optics tube' and will have unique lens shapes. 
Separating the lenses and filters into individual optics tubes, each housing an individual focal plane module, allows for improved aberration control and significantly smaller element sizes in comparison to SPT-3G.
For example, the first lens in \sptthreegplus{} will be made of alumina and have a clear aperture of 240~mm, whereas the first lens in \sptthreeg{} was also alumina but with a 720~mm aperture. 
Smaller lenses are easier to procure, handle, cool, apply robust broadband anti-reflection features, and thinner, with lower in-band absorption, resulting in significantly improved end-to-end optical efficiency.
As seen in Figure \ref{fig:optics_tube}, each 750~mm long optics tube will have four cryogenic lenses: one alumina lens; two silicon lenses; and one nylon lens. 
The lens materials were chosen due to their high mm-wavelength refractive index, relatively low-loss in-band, and their ability to act as effective infrared filters, in the case of alumina and nylon.
To reduce reflections, each cryogenic optical element will have metamaterial anti-reflection features machined into its surface \cite{golec20}. 

The \sptthreegplus{} optical system has a 4~degree diameter field of view with a Strehl ratio $>$0.9 across the entire focal plane. 
The angular resolution at 90 and 150~GHz will be 1.7 and 1.2 arcmin, respectively. 
For a more detailed discussion of the structure and elements within an optics tube see Athreya et al. 2026 in this publication preceding.
   
\subsection{Receiver Design}

\begin{figure} 
   \begin{center}
    \includegraphics[width=1.0\textwidth]{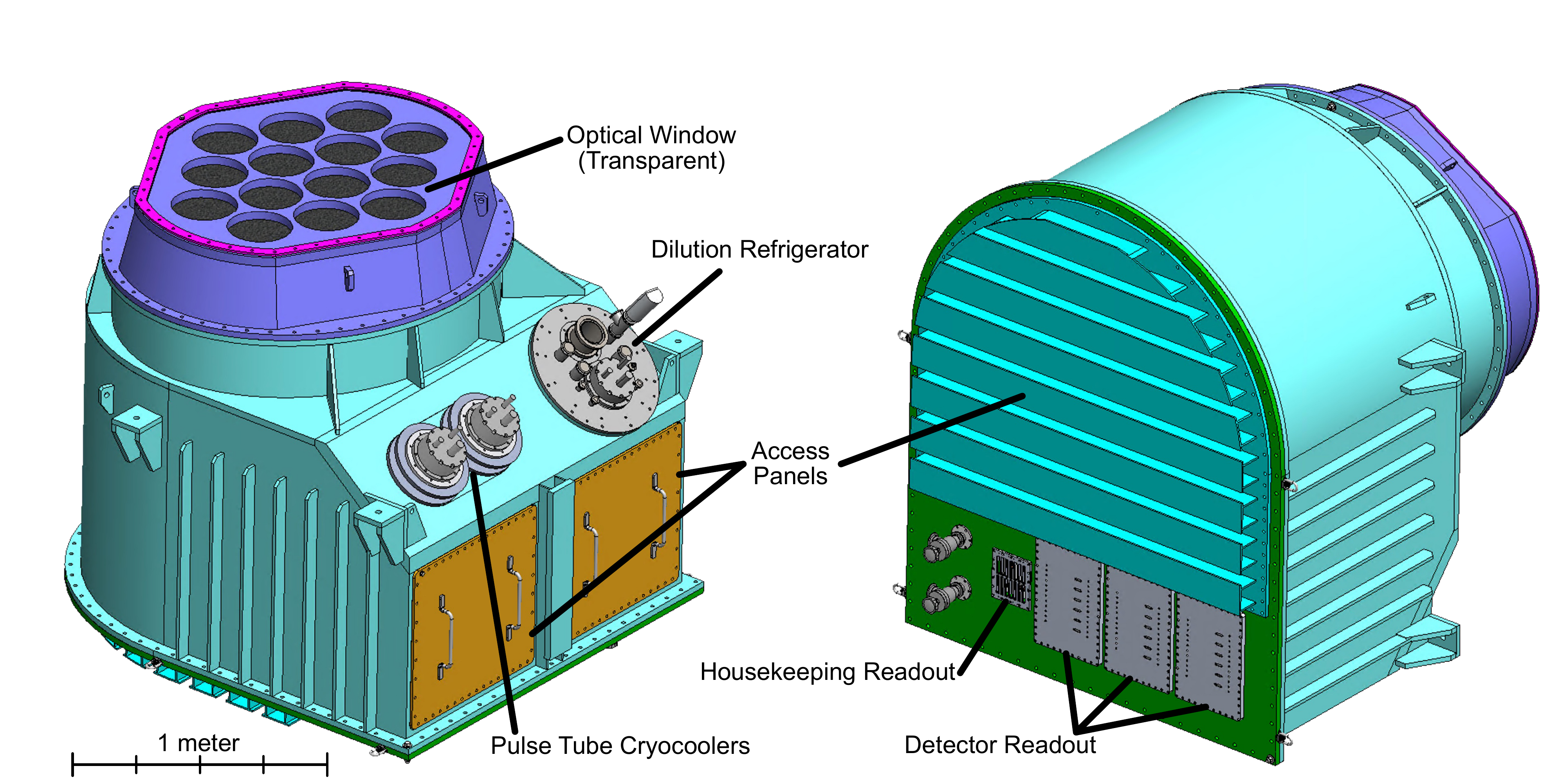}
   \end{center}
   \caption{ \label{fig:S3G+_cryo_iso_top} 
    Two views of the \sptthreegplus{} receiver are shown with the dilution refrigerator and pulse tube cryocoolers installed. 
    The side and bottom access panels are removed to install optic tubes and work on internal components. 
}
   \end{figure}

The \sptthreegplus{} receiver was designed to maximize the number of dichroic 90 and 150~GHz pixels.
This resulted a receiver design that will house the largest focal plane installed on the SPT at $\sim$1~m in diameter. 
The receiver cryostat will measure roughly 1.7x1.8x2~m with removable panels for access located on the bottom and side of the vacuum vessel (see Figure \ref{fig:S3G+_cryo_iso_top}). 
Because the \sptthreegplus{} receiver will be installed in a telescope cabin that was originally designed to accommodate the much smaller SPT-SZ receiver, space constraints within the cabin dictate some of the geometric designs of the receiver. 
The rounded end of the receiver and the pseudo-octagonal shape of the upward extension of the receiver are examples of designs made to fit the receiver within the telescope cabin and avoid structural components while still maximizing the potential focal plane area. 
The receiver cryostat shell requires a number of external fins and gussets to minimize the shell deflections, material stress, and vacuum integrity when evacuated.

The \sptthreegplus{} receiver will have inner radiation shells at $\sim$40 K and $\sim$4 K that will be cooled by a Cryomech pulse tube cryocooler (PTC) using helium gas. 
This PTC will also cool elements of the optics tubes that operate at $\ge$4~K. 
The receiver will have the ability to house and run two PT420 cryocoolers to mitigate the risk of a higher thermal load than anticipated, although only one cryocooler is expected to be required. 
The \sptthreegplus{} focal plane will be cooled to 100~mK.

While previous SPT receivers used a three-stage \textsuperscript{4}He-\textsuperscript{3}He-\textsuperscript{3}He system to cool their focal planes to $\sim$250~mK, the \sptthreegplus{} receiver will use an SD250 dilution refrigerator manufactured by BlueFors\footnote{bluefors.com} to achieve a 100~mK base temperature.
This dilution refrigerator will provide $>$250 $\mu$W of continuous cooling power at 100~mK. 
Operating at 100~mK compared to 250~mK lowers the detector noise and allows the use of microwave multiplexing readout (Section \ref{readout}).
The continuous cooling provided by the dilution refrigerator will allow for a $>15\%$ increase in observing efficiency compared to the $\sim$80$\%$ cooling duty cycle of the three stage cooling system used by \sptthreeg. 

For a more detailed discussion of the \sptthreegplus{} receiver design, please see Athreya et al. 2026 in this publication preceding.

\subsection{Focal Plane Modules}
\begin{figure} [ht]
   \begin{center}
   \begin{tabular}{c} 
   \includegraphics[height=10cm]{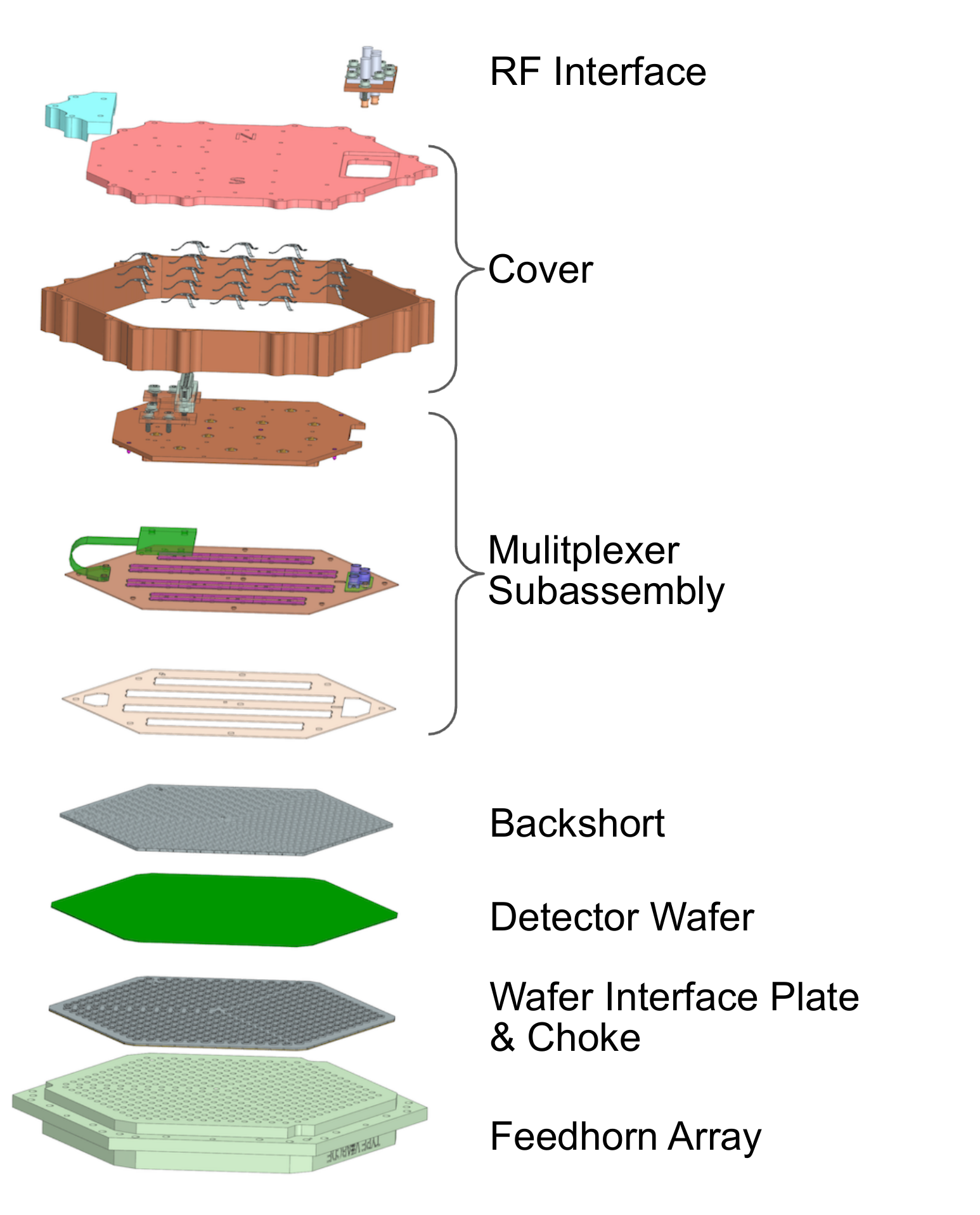}
   \end{tabular}
   \end{center}
   \caption[example] 
   { \label{fig:SPT3G+_Module_exploded} 
    Exploded view of a module identifying its many components. 
    }
   \end{figure}

A focal plane module consists of a feedhorn-coupled detector array and cryogenic multiplexer, packaged into a compact hexagonal geometry.  
The design follows the Simons Observatory implementation \cite{healy20,healy23,mccarrick21} but with some distinctions that we discuss below.   
Each of the 430 spatial pixels per module measure power in orthogonal polarizations in two wide bands centered near 90~GHz and 150~GHz.  
An exploded view of the module components is shown in Fig.~\ref{fig:SPT3G+_Module_exploded}.  
A prototype of a full focal plane module is currently being produced.
In the subsections below, we describe the detector arrays, multiplexed readout, and mechanical packaging aspects in more detail.

\subsubsection{Detector Wafers}

\sptthreegplus{} uses multichroic transition-edge sensor (TES) bolometer arrays coupled to planar ortho-mode transducers (OMTs) and feedhorns to enable polarimetric imaging in frequency bands centered near 90~GHz and 150~GHz.  
Detectors of this architecture have been deployed in a number of CMB experiments \cite{datta16,henderson16,galitzki18}.
The OMT-coupled TES circuit was first described in McMahon et al. \cite{mcmahon12}, the millimeter-wave component design is presented in Hubmayr et al. \cite{hubmayr22}, and fabrication details are discussed in Duff et al.\cite{duff16} and Cecil et al. \cite{Cecil2020}.
In brief, polarization-sensitive niobium (Nb) probes on a thin silicon nitride membrane located within a circular waveguide at the exit of the feedhorn launch signal onto a planar superconducting circuit.
On-chip diplexers divide the signal into two wide frequency bands before each divided signal passes through stub filters that define the detector bandpasses. 
The divided and filtered signals from opposing Nb probe arms are combined, and single-mode power within each frequency band and in each polarization is detected by one of four voltage-biased TES bolometers.  
The \sptthreegplus{} wafer design draws from Advanced ACTPol, Simons Observatory, and CMB-S4 mid-frequency (MF) designs \cite{henderson16,duff24,barron22}.  
The well-developed microfabrication process \cite{duff26} has led to 38 science grade Simons Observatory MF wafers exhibiting detector properties that match specifications \cite{dutcher24,abitbol25}. 
Relative to these past implementations, the bolometer properties tabulated in Tab.~\ref{tab:detector_parameters} have been optimized for the loading conditions of SPT.  
We aim to achieve the bolometer saturation power target by geometric scaling of the thermally isolating bolometer legs. 
Matching previous designs, the TESs are composed of AlMn films \cite{li16} with target superconducting critical temperature T$_{\rm{c}}$~=~160~mK and normal resistance R$_{\rm{n}}$~=~8~m$\Omega$.  

 \begin{figure}[htbp]
    \centering
    \begin{minipage}{0.48\textwidth}
        \centering
        \includegraphics[width=\linewidth]{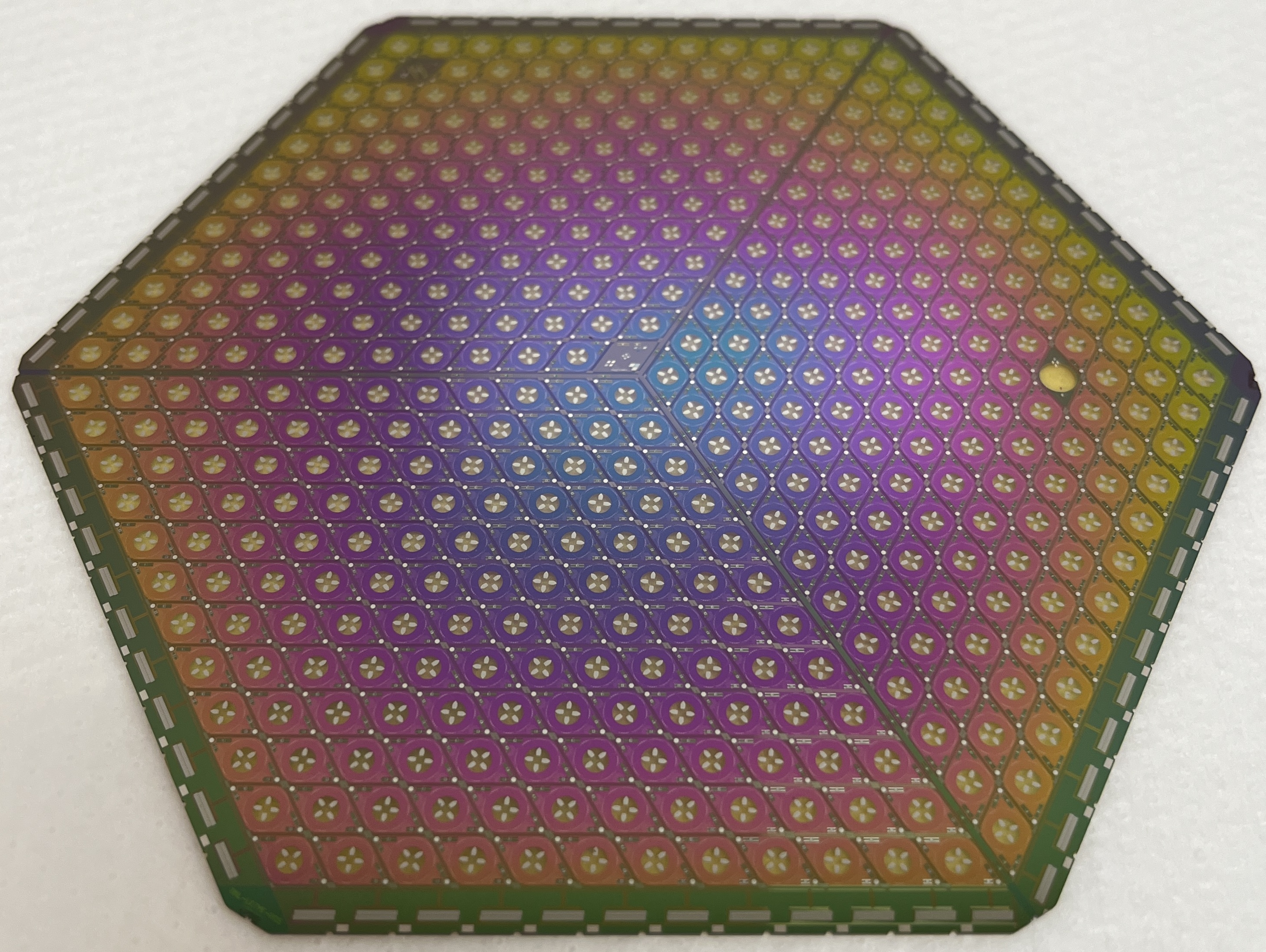}
        \label{fig:sub1}
    \end{minipage}
    \hfill
    \begin{minipage}{0.48\textwidth}
        \centering
        \includegraphics[width=\linewidth]{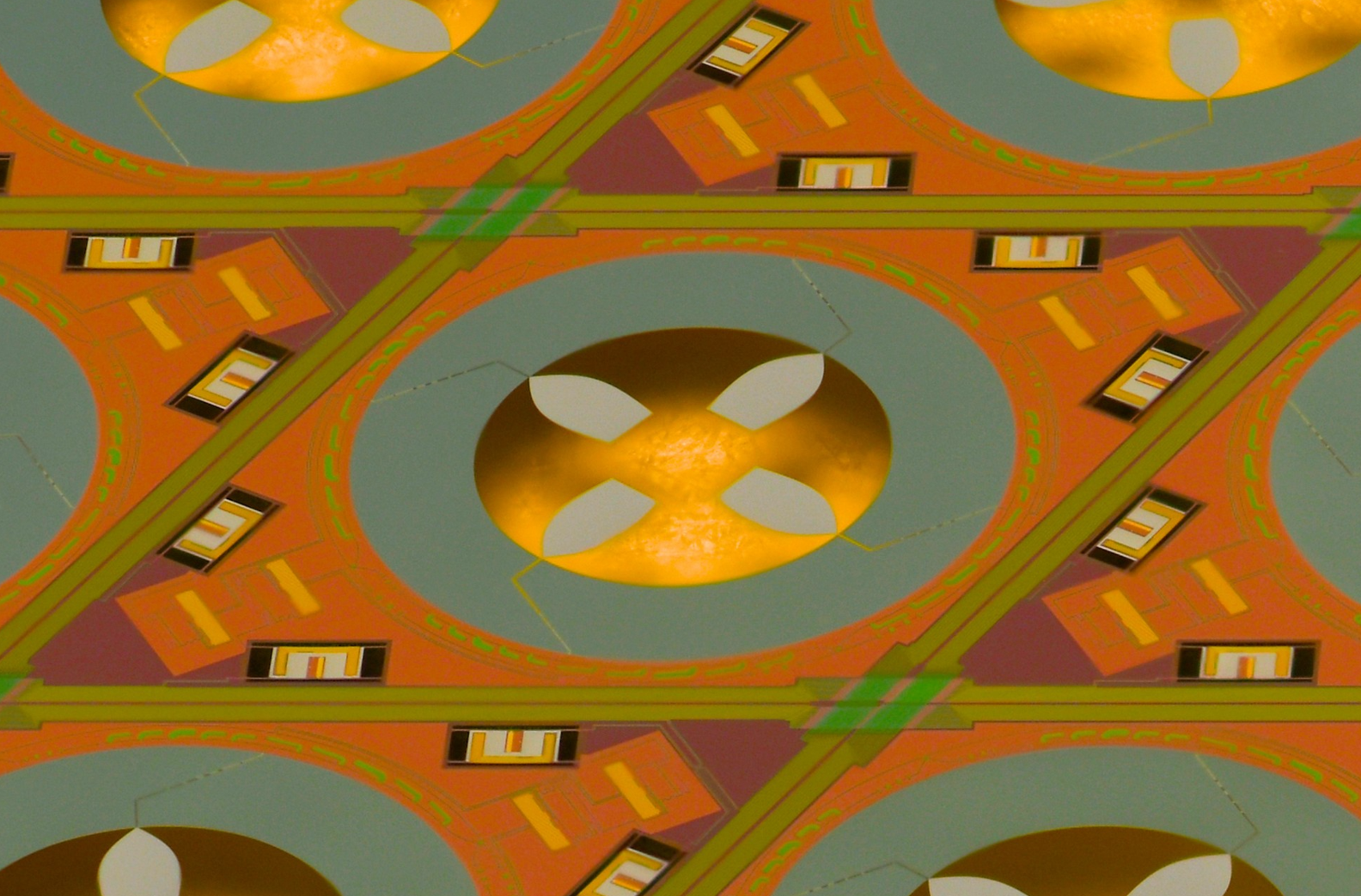}
        \label{fig:sub2}
    \end{minipage}
    \vspace{0.5em}
    
    \begin{minipage}{0.48\textwidth}
        \centering
        \includegraphics[width=\linewidth]{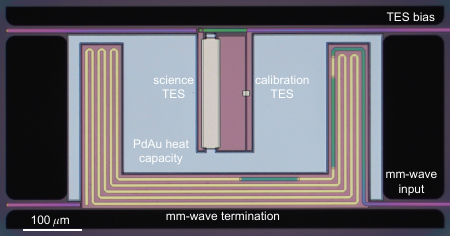}
        \label{fig:sub3}
    \end{minipage}
    \hfill
    \begin{minipage}{0.48\textwidth}
        \centering
        \includegraphics[width=\linewidth]{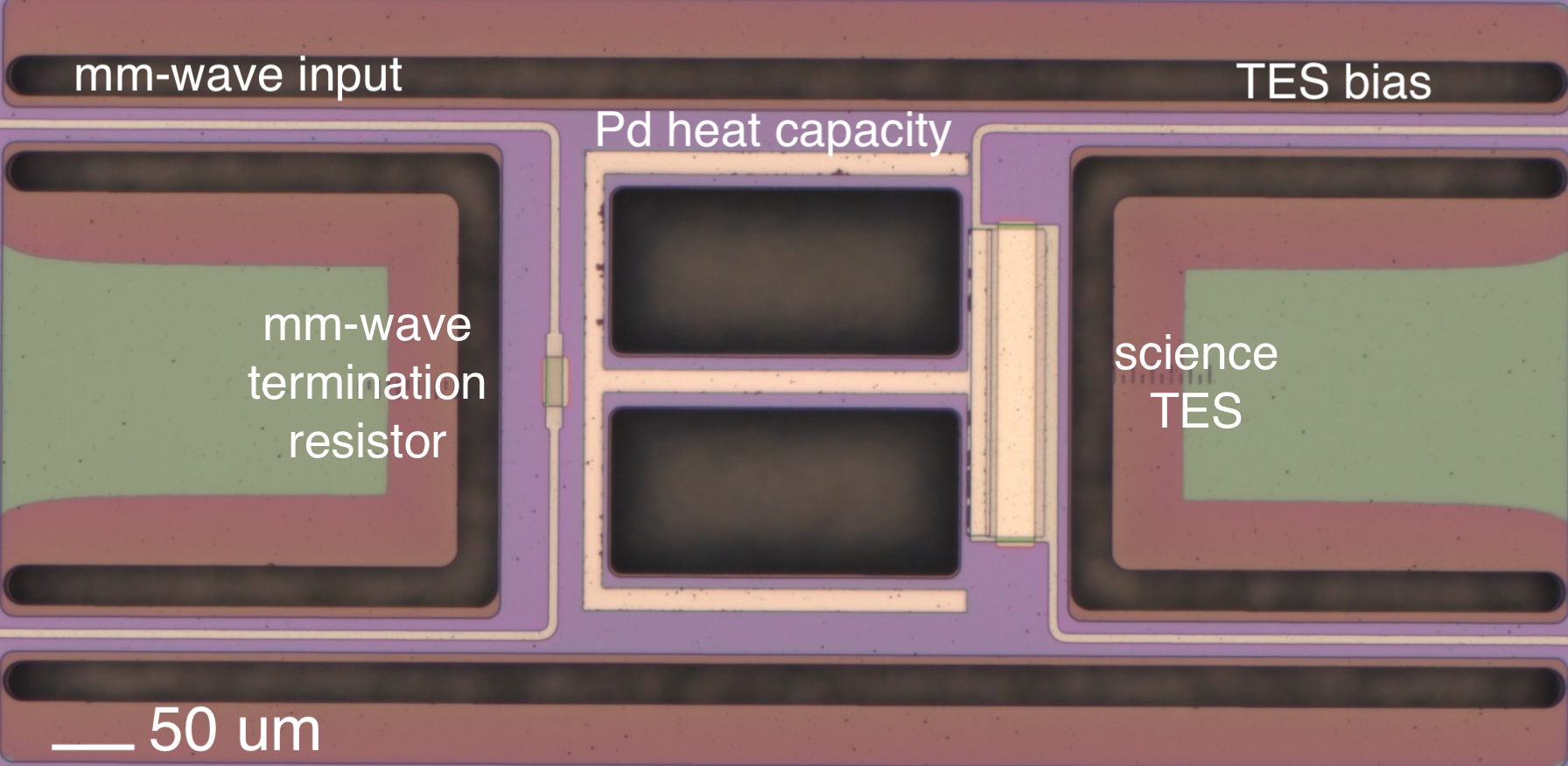}
        \label{fig:sub4}
    \end{minipage}
    \caption{Examples of prototype detector wafers, moving clockwise from upper left: A full 150~mm ANL wafer is shown with 430 detector pixels;  a single pixel from a NIST wafer; a bolometer from ANL with the TES on the right and the lumped termination on the left; a bolometer from NIST with the 'science' TES on the center left, the 'calibration' TES on the center right, and the lossy meander at the bottom and sides.}
    \label{fig:composite}
\end{figure}

The 14 wafers will be fabricated at two different fabrication locations, Argonne National Laboratory (ANL) and the National Institute of Standards and Technology (NIST). 
The use of two fabrication locations provides redundancy, reducing risk and shortening the instrument build timeline.   
While detector specifications are identical, implementation at each fabrication location matches the microfabrication toolset and expertise.  
Wafers from each fabrication site differ in two main respects: how signal from opposing Nb OMT fins are combined and how bolometers are released from the substrate.  
NIST uses hybrid-Ts and a lossy meander to combine and terminate the signal, whereas ANL implements a symmetrically fed lumped termination resistor located on the bolometer island.  
To release the bolometers, NIST uses a deep reactive ion etch (DRIE) step, which also defines the OMT membrane in the same fabrication step.  
In contrast, ANL uses a XeF2 etch to release the bolometer islands and DRIE to define the OMT membrane in separate fabrication steps.  
Figure~\ref{fig:composite} shows images of wafer prototypes fabricated at NIST and ANL, which highlights these distinctions.  
The prototype wafer from NIST includes a `science' TES in series with a proximity effect `calibration' TES \cite{sadleir10,nagler20}, which can be used in a laboratory setting with higher loading. 

\begin{table}[htbp]
    \centering
    \caption{Detector Parameter Targets}
    \label{tab:detector_parameters}
    \begin{tabular}{|c|c|c|}
        \hline
        \textbf{Parameter} & \textbf{90 GHz} & \textbf{150 GHz}\\
        \hline
        $P_{sat}$ (pW)       &  6.1 & 12.5 \\
        Band center (GHz)    &  91.5 & 148.5 \\ 
        Bandwidth (GHz)      &  29 & 42 \\ 
        $R_{tes}$ (mOhms)    &  8 & 8 \\
        $T_c$ (K)            &  0.16 & 0.16 \\
        \hline
    \end{tabular}
\end{table}

\subsubsection{Readout}
\label{readout}

Detectors are read out with a microwave SQUID multiplexer system (\umux), which was specifically developed to enable the use of large arrays of TES bolometers \cite{irwin04,mates11,dober21}.  
In this system, TES bolometers are DC biased and the current flowing through each TES is flux-coupled to a radio-frequency Superconducting Quantum Interference Device (SQUID) which terminates a microwave resonator. 
The changing TES current causes a frequency shift in the microwave resonator, which can be measured with standard homodyne readout techniques. 
Since each resonator has a unique frequency, many detector signals can be read out, or multiplexed, over a single coaxial cable.
Following developments from the Simons Observatory, the 1756 TES bolometers in a \sptthreegplus{} detector wafer (1720 optical and 36 dark) will be read out by two 910-channel cryogenic multiplexers in the 4-6~GHz readout band using the 300~K SMuRF electronics developed by SLAC \cite{henderson18,yu22,yu23}.
While multiplexing factors up to 1820 have been demonstrated \cite{groh25}, for \sptthreegplus{} the benefits of reducing the number of cables and 4~K amplifiers by a factor of two do not outweigh the technical maturity of the 2$\times$910 multiplexer approach.   

With these choices, the 24,080 optical and 364 dark detectors in the \sptthreegplus{} focal plane will be read out using 28 coaxial cables. 
There will also be 196 twisted-wire pairs to provide the detector bias (12 pairs per module) and \umux flux ramp modulation (2 pairs per module) \cite{mates11}. 
    
\subsubsection{Module Packaging}
The \sptthreegplus{} modules build off of the Simons Observatory design, which integrates the optical coupling, detector wafers, and readout into one package shown in Figure \ref{fig:SPT3G+_Module_exploded} \cite{mccarrick21, healy23}. 
A spline-profiled gold-coated Al feedhorn array \cite{simon18} defines the detector beam and is coupled to the detector stack with three Si optical coupling wafers: the choke, the waveguide interface plate (WIP), and the backshort. 
The choke reduces the leakage at the interface between the horn array and Si optical coupling wafers, and the WIP is coupled to the detector wafer to enable continuity in the waveguide. 
The backshort is added behind the detector wafer and is optimized to maximize the signal coupled to the detectors. 
The components are kept in alignment as they cool by a pin and slot system. 
The detector wafer is bonded to the feedhorn array with Au wire bonds for heat sinking.  

The optical coupling and detector wafers are integrated with the 2$\times$910 multiplexer sub-assembly.  
This subassembly contains 28 multiplexer chips \cite{dober21}, a silicon routing wafer \cite{duff26} that includes passive detector bias components and distributes TES leads to the SQUID readout, and a precision machined metal tray and lid.  
The 500~$\mu$m thick tray base allows for short `step-up' wire bonds from the detector wafer to the routing wafer, which enables the placement of all readout elements behind the detectors.  
In contrast to the SO design and potentially facilitating the machining, we will evaluate the use of an aluminum tray and lid (as opposed to copper) in the first prototype that is currently being assembled.  
The microwave SQUIDs are known to be sensitive to stray magnetic fields \cite{connors22, vavagiakis21b, huber22}.  
By surrounding the SQUIDs with superconducting packaging, both the magnetic field and heat sinking environment will be substantially altered.  
Upcoming measurements, in which the full detector package will be carefully degaussed, will determine if this change is beneficial and set the final tray material for the production modules. 

Additional changes from the SO design are for convenience.  
We use a high density connector on the \umux\ tray and commercial BeCu springs to press the multiplexer and detector sub-assemblies onto the feedhorn array.  
The estimated mass of the full module, using all aluminum, is 1.5~kg.     

\section{Summary}

In this work, we have described the design and science goals of \sptthreegplus{}, which will be the next survey instrument installed on the SPT and targeting to start observations in early 2029.
The \sptthreegplus{} instrument will improve the field of view, detector packing density, optical efficiency, and readout noise compared to \sptthreeg.
These improvements will allow \sptthreegplus{} to achieve a CMB mapping speed nearly an order of magnitude larger than \sptthreeg{}. 
The \sptthreegplus{} data will make unprecedentedly deep, arcminute angular resolution maps of the CMB, that will be used to enable new studies of cluster formation and cosmology; make high-cadence, high-sensitivity observations of the time-variable mm-wavelength sky to study transient sources; and produce the highest signal-to-noise lensing map yet from the CMB.  
This lensing map will be used to delens foreground B modes and in combination with BICEP, as part of the SPO, aims to achieve a measurement of $\sigma(r) = 0.001$.

\appendix    

\acknowledgments 
The South Pole Telescope program is supported by the National Science Foundation (NSF) through awards OPP-2332483 and OPP-2408494.
Argonne National Laboratory's work was supported by the U.S. Department of Energy, Office of High Energy Physics, under contract DE-AC02-06CH11357.
Work performed at the Center for Nanoscale Materials, a U.S. Department of Energy Office of Science User Facility, was supported by the U.S. DOE, Office of Basic Energy Sciences, under Contract No. DE-AC02-06CH11357.
This document was prepared by the SPT-3G+ collaboration using the resources of the Fermi National Accelerator Laboratory (Fermilab), a U.S. Department of Energy, Office of Science, Office of High Energy Physics HEP User Facility. Fermilab is managed by Fermi Forward Discovery Group, LLC, acting under Contract No. 89243024CSC000002.
The SLAC group is supported in part by the Department of Energy at SLAC National Accelerator Laboratory, under contract DE-AC02-76SF00515.
This research was funded by the National Institute of Standards and Technology (ror.org/05xpvk416) and the University of Chicago (ror.org/024mw5h28) under agreement number 300004034.
This work is supported by the Gordon and Betty Moore Foundation, through Award \#14367.
The work at Case Western Reserve University is partially supported by the U.S. Department of Energy under award number DE-SC0009946.
Work at the University of Illinois Urbana-Champaign is partially supported by the U.S. Department of Energy under award number DE-SC0015655.
D.R.B. and W.D. were supported by DOE HEP under award DE-SC0021435.
The Melbourne authors acknowledge support from the Australian Research Council's Discovery Project scheme (No. DP210102386). 
The Paris group has received funding from the European Research Council (ERC) under the European Union's Horizon 2020 research and innovation program (grant agreement No 101001897), and funding from the Centre National d'Etudes Spatiales. 
R. A. Lew's work was performed under financial assistance award 70NANB21H182 and 70NANB25H091 from the National Institute of Standards and Technology, U.S. Department of Commerce. The statements, findings, conclusions, and recommendations are those of R. A. Lew and do not necessarily reflect the views of the National Institute of Standards and Technology or the U.S. Department of Commerce.
H. Athreya's work is partially supported by the University of Chicago Physical Sciences Division's Eckhardt Graduate Scholarship.

\bibliography{BIBTEX/spt_1995_to_2000,BIBTEX/spt_2000_to_2005,BIBTEX/spt_2005_to_2010,BIBTEX/spt_2010_to_2015,BIBTEX/spt_2015_to_2020,BIBTEX/spt_2020_to_2025,BIBTEX/spt_2025_and_after,BIBTEX/spt_before_1995,BIBTEX/3G+_natoli} 
\bibliographystyle{spiebib} 

\end{document}